\documentstyle[aps,preprint]{revtex}
\begin{document}
\draft
\tightenlines
 
\title{Quasi-stationary simulation of the contact process}
\author{Ronald Dickman$^\dagger$ and Marcelo Martins de Oliveira$^*$ 
}
\address{
Departamento de F\'{\i}sica, ICEx,
Universidade Federal de Minas Gerais,\\
30123-970
Belo Horizonte - Minas Gerais, Brasil\\
}

\date{\today}

\maketitle
\begin{abstract}
We review a recently devised Monte Carlo simulation method
for the direct study of quasi-stationary properties of 
stochastic processes with an absorbing state.  The method is
used to determine the static correlation function and the
interparticle gap-length distribution in
the critical one-dimensional contact process.
We also find evidence for power-law decay of the interparticle
distance distribution in the two-particle subspace. 
\end{abstract}


\noindent {\small $^\dagger$electronic address: dickman@fisica.ufmg.br}

\noindent {\small $^*$electronic address: mancebo@fisica.ufmg.br}

\newpage
\section{Introduction}

Stochastic processes with an absorbing state arise frequently in
statistical physics \cite{vankampen,gardiner}, epidemiology
\cite{bartlett} and related fields.
Phase transitions to an absorbing state in spatially extended
systems, exemplified by the contact process \cite{harris,liggett}, 
are currently of great interest
in connection with self-organized criticality \cite{socbjp}, 
the transition to turbulence \cite{bohr}, and 
issues of universality in nonequilibrium
critical phenomena \cite{marro,hinrichsen,odor04}.  

The {\it quasi-stationary} (QS) distribution, (that is, conditioned
on survival), is very useful in the study of
processes with an absorbing state model.
Conventional simulations of ``stationary" properties of lattice 
models with an absorbing state actually study the quasi-stationary regime,
given that the only true stationary state for a finite system is the absorbing 
one.  
We recently devised a simulation 
method that yields quasi-stationary properties directly \cite{qssim};
it samples the QS 
probability distribution
just as conventional Monte Carlo simulation 
samples the equilibrium distribution. Here we use the method  
to study the static correlation function 
and other configurational properties of the critical contact process,
the prime example of an absorbing-state phase transition.

In the following section we review the basis of our method.
Then in Sec. III we apply it to determine the static two-point correlation
function of the contact process on a ring.  We summarize our findings in Sec. IV.

\section{Background}

Consider a continuous-time Markov process $X_t$ taking values 
$n = 0, 1, 2,...,S$, 
with state $n\!=\!0$ absorbing.  
We use $p_n(t)$ to denote the probability that $X_t = n$, given some initial
state $X_0$. The survival probability is
 $P_s(t) = \sum_{n \geq 1} p_n(t) = 1 - p_0(t)$.
We suppose that as $t \to \infty$ the $p_n$, normalized by
the survival probability $P_s(t)$, attain a time-independent 
form, thus defining the
quasi-stationary distribution $\overline{p}_n$: 

\begin{equation}
\overline{p}_n  \equiv \lim_{t \to \infty} \frac{p_n (t) }{P_s (t)} 
,\;\;\;\; (n \geq 1),
\label{qshyp}
\end{equation}
with $\overline{p}_0 \equiv 0$. (We further assume that the limiting distribution
does not depend on the initial state, as long as $X_0 \neq 0$.)The QS distribution is
normalized so:
\begin{equation}
\sum_{n \geq 1} \overline{p}_n = 1.
\label{norm}
\end{equation}

As shown in \cite{RDVIDIGAL}, the QS distribution is the 
stationary solution to
the following equation of motion  (for $n > 0$)

\begin{equation}
\frac{d q_n}{dt} = -w_n q_n + r_n + r_0 q_n \;,
\label{qme}
\end{equation}
where $w_n = \sum_m w_{m,n}$ is the total rate of transitions out of state
$n$, and $r_n = \sum_m w_{n,m} q_m$ is the flux of probability into this
state.  To see this, consider the master equation (Eq. (\ref{qme})
without the final term) in the QS regime.  Substituting
$q_n(t) = P_s(t) \overline{p}_n$, and noting that in the QS regime
$d P_s/dt = -\overline{r}_0 = - P_s \sum_m w_{0,m} \overline{p}_m$, 
we see that the r.h.s. of Eq. (\ref{qme}) is identically zero
if $q_n = \overline{p}_n$ for $n \geq 1$.
The final term in Eq. (\ref{qme}) represents a redistribution of the 
probability $r_0$ (transfered to the
absorbing state in the original master equation), 
to the nonabsorbing subspace.
Each nonabsorbing state receives a share
equal to its QS probability.  

In \cite{qssim} we introduce a process $X_t^*$, whose {\it stationary} 
probability distribution is the {\it quasi-stationary} distribution 
of $X_t$.
(Note that in order to have a nontrivial stationary
distribution, $X_t^*$ cannot
possess an absorbing state.)  
The probability distribution of $X_t^*$ is governed by Eq. (\ref{qme}),
which implies that for $n>0$ (i.e., away from the absorbing
state), the evolution of $X_t^*$ is identical to that of $X_t$.
When $X_t$ enters the absorbing state, however, $X_t^*$ instead
jumps to a nonabsorbing one, and then resumes its ``usual" evolution
(with the same transition probabilities as $X_t$), until such time
as another visit to the absorbing state is imminent.

In Eq. (\ref{qme}) 
the distribution $q_n$ is used to determine the value of
$X_t^*$ when $X_t$ visits the absorbing state.  Although one
has no prior knowledge of $q_n$ (or its long-time limit,
the QS distribution $\overline{p}_n$),
one can, in a simulation, use the 
history $X_s^*$ ($0 < s \leq t$) up to time $t$, to
{\it estimate} the $q_n$.  
This is done by saving 
a sample $\{n_1, n_2, ..., n_M\}$ of configurations visited.
We update the sample  
by replacing from time to time one of the configurations 
with the current one.  
In this way the distribution for the process $X_t^*$ 
will converge to the QS distribution (i.e., the stationary
solution of Eq. (\ref{qme})) at long times.
Summarizing, the process $X_t^*$ has the same dynamics as 
$X_t$, except when a transition to the absorbing 
state is imminent: $X_t^*$ then is placed in a nonabsorbing state, 
selected at random from a sample over the history of the realization.
(In the simulation, a list of $M$ configurations is maintained.  Whenever
the time increases by 1, the list is updated with probability $p_{rep}$,
by replacing a randomly chosen configuration on the list with the current one.)
In effect, the final term in Eq. (\ref{qme}) is represented as a 
{\it memory} in the simulation.

The above scheme was shown \cite{qssim}
to yield precise results, in accord with the
exact QS distribution for the contact process on a complete graph
\cite{RDVIDIGAL}, and with conventional simulations of the same model
on a ring, for which exact results are not available.
QS simulation results were found rather insensitive to the choice
of list size $M$ and replacement rate $p_{rep}$.  In the studies 
reported below we use $M=1000$ and $p_{rep} = 0.001$.

\section{Contact Process: Correlation Function}

The contact process (CP) \cite{harris} is a continuous-time
Markov process on a lattice, in which
each site $i$ is 
either occupied ($\sigma_i (t)= 1$),
or vacant ($\sigma_i (t)= 0$).  Transitions from $\sigma_i = 1$ to 
$\sigma_i = 0$ occur at a rate of unity, independent of the neighboring sites.  
The reverse transition can only occur if at least one neighbor is 
occupied: the transition from $\sigma_i = 0$ to $\sigma_i = 1$ 
occurs at rate 
$\lambda r$, where $r$ is the fraction of nearest neighbors of site $i$ 
that are occupied; thus the state $\sigma_i = 0$ for all $i$ is absorbing.
($\lambda $ is a control parameter governing the rate of spread of
activity.)
The order parameter 
$\rho = \langle \sigma_i \rangle$ is the fraction of occupied sites.
The model exhibits a continuous phase transition at 
$\lambda_c = 3.297848(20)$ \cite{iwanrd93}.  For $\lambda < \lambda_c$,
the stationary value of $\rho$ is zero.

The CP has attracted much interest as a 
prototype of a nonequilibrium critical point, a simple representative of the
directed percolation (DP) universality class.  Since its scaling properties
have been discussed extensively \cite{marro,hinrichsen,odor04} we review
them only briefly.  As the critical point is approached, the
correlation length $\xi$ and correlation time $\tau$ diverge,
following $\xi \propto |\Delta|^{-\nu_\perp}$ and 
$\tau \propto |\Delta|^{-\nu_{||}}$, where
$\Delta = \lambda - \lambda_c$ is the distance from the critical point.  The order 
parameter
scales as $\rho \propto \Delta^\beta$ for $\Delta > 0$.
At the critical point the quasi-stationary value of the
order parameter scales as: $\rho \propto L^{-\beta/\nu_\perp}$.

An aspect of the CP that has not, to our knowledge, been studied in
simulations is the static correlation function.  Of particular interest is how
correlations decay at the critical point.  The correlation function is defined via
\begin{equation}
C(|i\!-\!j|) = \langle \sigma_i \sigma_j \rangle - \langle \sigma_i \rangle  \langle\sigma_j \rangle
\label{defCr}
\end{equation}
where the average is over the stationary distribution of the process.  

Based on experience with equilibrium critical phenomena, we expect
$C(r)$ to decay as a power law at the critical point.  The decay exponent
can be determined via a scaling argument.  To begin, we normalize
$C(r)$ to its value at $r=0$:
\begin{equation}
c(r) \equiv \frac {C(r)}{C(0)} = \frac{C(r)}{\rho (1\!-\!\rho)}
\label{defcr}
\end{equation}
Now consider the scaled variance
\begin{equation}
\chi = L^d \left( \langle \rho^2 \rangle - \rho ^2 \right)
\label{chi}
\end{equation}
($L$ denotes the lattice size.)  At the critical point,
$\chi \sim L^{\gamma/\nu_\perp}$ with $\gamma = d \nu_\perp - 2 \beta$
\cite{marro}.  A simple calculation yields
\begin{equation}
\chi = L^{-d} \sum_{i,j} C(|i\!-\!j|) 
= \rho (1-\rho) \sum_{|\bf{x}| \leq {\it L/2}} c(|\bf{x}|)
\label{chiCr}
\end{equation}
where in the last step we used translation invariance, and assumed a
hypercubic lattice of $L^d$ sites.  Now suppose $c(r) \sim r^{-\alpha}$
for large $r$.  Approximating the sum by an integral, and recalling that
$\rho \sim L^{-\beta/\nu_\perp}$ at the critical point, we find
\begin{equation}
\chi \sim L^{d - 2 \beta/\nu_\perp} \sim L^{d - \alpha - \beta/\nu_\perp}
\label{scalalpha}
\end{equation}
implying that $\alpha = \beta/\nu_\perp$
and 
\begin{equation}
C(r) \simeq \rho c(r) \propto (rL)^{-\beta/\nu_\perp} 
\label{scalC}
\end{equation}
in the critical stationary state.  The relation $\alpha = \beta/\nu_\perp$
was demonstrated some time ago by Grassberger and de la Torre, who showed
that $C(r) \sim r^{-2 \delta/z}$ at the critical point \cite{torre}.
(Their result is seen to be equivalent to ours when we recall the
scaling relations $z = 2\nu_\perp/\nu_{||}$ and $\delta = \beta/\nu_{||}$.)

We use the QS simulation method outlined above to determine
the correlation function.
The process is simulated in five independent realizations 
of $2 \times 10^8$ time steps.
As is usual, annihilation events are chosen with probability $1/(1+\lambda)$ and
creation with probability $\lambda/(1+\lambda)$.  A site is chosen from
a list of currently occupied sites, and, in the case of annihilation, is 
vacated, while, for creation events, a nearest-neighbor site is selected 
at random and, if it is currently vacant, it becomes occupied.  The time 
increment associated with each event is $\Delta t = 1/N_{occ}$, where 
$N_{occ}$ is the number of occupied sites just prior
to the attempted transition \cite{marro}.

In Fig. 1 we plot $C^* (r) = L^{\beta/\nu_\perp} C(r)$ for 
$L=1280$ and 2560, using
the  best available estimate (from series analysis),
$\beta/\nu_\perp = 0.252072(8)$ \cite{jensen}.
The data collapse for the two lattice sizes is nearly perfect.
For $r \ll L$ the correlation function indeed follows a power law
$C^* \sim r^{-\beta/\nu_\perp}$, while for $r=L/2$ it attains a
minimum, as expected due to the periodic boundaries.  To determine
the decay exponent we analyze the local slope $\alpha(r)$ (see Fig. 2), 
obtained from a linear fit to the data for $\ln C^*$ versus $\ln r$, using
points equally spaced in $\ln r$, in finite intervals $[r_0, 3r_0]$.
For $r_0 \ll L$ the local slope is nearly constant, but it of course
veers upward as $r_0$ approaches $L/2$.  We therefore perform an 
extrapolation (to $r \to \infty$) of the local slope versus $1/r$,
using only the data on which the results for the two lattice sizes 
agree, to eliminate finite-size effects.  The result of this extrapolation
is $\beta/\nu_\perp = 0.252(1)$, consistent with the best
estimate.

For $\lambda < \lambda_c$, the correlation function decays exponentially,
$C(r) \sim e^{-r/\xi}$.  This is evident in the inset of
Fig. 1, where we plot $\tilde{C} = r^{\beta/\nu_\perp} C(r)$.  Exponential
decay is clear for $\lambda = 0.99 \lambda_c$; linear regression 
yields $\xi \simeq 356$ in this case, well below the system
size ($L=2560$) in this study.  Interestingly, the
decay of $\tilde{C}$ is also perceptible for $\lambda = 0.999 \lambda_c$,
though partly masked due to the finite system size.
In general the decay of correlations should be evident for $L > \xi$,
where $\xi$ is the correlation length in the infinite-size limit.
Since $\xi \sim |\Delta|^{-\nu_\perp}$, deviations from criticality
on the order of $|\Delta| \sim L^{-1/\nu_\perp}$ (or greater)
should be detectable in the correlation function for a system of size $L$.

In the critical stationary state, the distribution of particles is scale-invariant,
as reflected in the power-law decay of $C(r)$.  The distribution of {\it gaps}, or
strings of empty sites between successive occupied sites also follows a power law.
(A gap of size $g$ corresponds to sites $i$ and $i+g+1$ occupied and all intervening sites
empty.)  We determine the gap-size distribution $P(g)$ (normalized to the number of
gaps in the configuration).
As shown in Fig. 3, $P(g)$ exhibits a power law decay, $P \sim g^{-\tau}$, 
with $\tau \simeq 1.70$, over an intermediate range that appears to grow with system size.
Analysis of the local slope yields $\tau = 1.73(1)$.

We can relate the exponent $\tau$ to other critical exponents via a
simple scaling argument.  Since there is one gap per particle,
the mean gap size $\langle g \rangle$ is just the reciprocal
of the particle density.  Thus in a system of size $L$ at the critical point,
$\langle g \rangle \sim L^{\beta/\nu_\perp}$.  Assuming $P(g) \sim g^{-\tau}$
for $g \geq 1$, we have
\begin{equation}
\langle g \rangle \sim \int_1^l g^{1-\tau} dg \sim L^{2-\tau} \sim 
L^{\beta/\nu_\perp}
\label{tauesc}
\end{equation}
implying $\tau = 2 - \beta/\nu_\perp \simeq 1.748$.  Our simulation result is
about 1\% smaller than the value predicted by the scaling argument.  
The discrepancy is likely caused by finite-size corrections that limit the
power-law regime of $P(g)$.

Finally, we report preliminary results
on a surprising behavior of the interparticle distance
in the two-particle subspace.  Let $d$ denote the separation between
the occupied sites, given that that exactly two sites are occupied.
(We take the minimum distance under periodic boundaries, 
so that $d \leq L/2$.)  Since particles are highly clustered 
in the critical CP, we should expect the two-particle
distance distribution $P_2(d)$ to decay with separation.  Our results (see Fig. 4)
from QS simulations at the critical point suggest a power-law decay,  
$P_2(d) \sim d^{-\kappa}$, with $\kappa \simeq 2.45$.  We have no
way of relating this exponent to the known critical exponents.  
Indeed, whether $P_2(d)$ follows a power-law will have to be confirmed
in larger-scale simulations.  This is somewhat challenging since,
as the system size increases, the probability of having exactly
two particles becomes ever smaller.

\section{Summary}

We have applied a new simulation method for
quasi-stationary properties to determine the
static correlation function $C(r)$ of the critical contact
process in one dimension.  Our results support the
behavior $C(r) \sim 1/(rL)^{\beta/\nu_\perp}$
anticipated from scaling arguments.
We also studied the gap-size distribution, which shows
evidence of a power-law decay, $P(g) \sim g^{-\tau}$,
with $\tau \simeq 2 - \beta/\nu_\perp$, the value predicted
by scaling.  Finally we note an apparent scale-invariant
behavior of the interparticle distance distribution
$P_2(d)$ in the two-particle subspace.
Study of the correlation function and the gap-size distribution
promise to be useful in characterizing scaling behavior of
new models, and may also be useful in locating the critical
point.

\vspace{1em}

\noindent{\bf Acknowledgment}

This work was supported by CNPq and FAPEMIG, Brazil.

\newpage

FIGURE CAPTIONS
\vspace{1em}

\noindent FIG. 1. 
QS simulation results for the scaled correlation function
$C^* = L^{\beta/\nu_\perp} C(r)$ in the critical one-dimensional
contact process.  Symbols: $\times$: $L=1280$; $+$: $L=2560$.
The slope of the straight line is -0.252.
Inset: semi-logarithmic plot of $\tilde{C} = r^{\beta/\nu_\perp}C(r)$
versus $r$ in a system of 2560 sites, for $\lambda = \lambda_c$ (upper curve), 0.1\% below $\lambda_c$
(middle) and 1\% below $\lambda_c$ (lower).
\vspace{1em}

\noindent FIG. 2.
Local slope $- \alpha (r)$ of the correlation function versus $1/r$.
Open symbols: $L=1280$; filled symbols: $L=2560$.
\vspace{1em}

\noindent FIG. 3. 
Gap-length distribution in the critical one-dimensional CP.  $\Box$:
$L=640$; $+$: $L=5120$.  The slope of the straight line is -1.73.
\vspace{1em}

\noindent FIG. 4. 
Distribution of interparticle distances $d$ in the two-particle
subspace of the critical CP. Open symbols: $L=640$; filled symbols: $L=1280$.
The slope of the straight line is -2.45.
\vspace{1em}

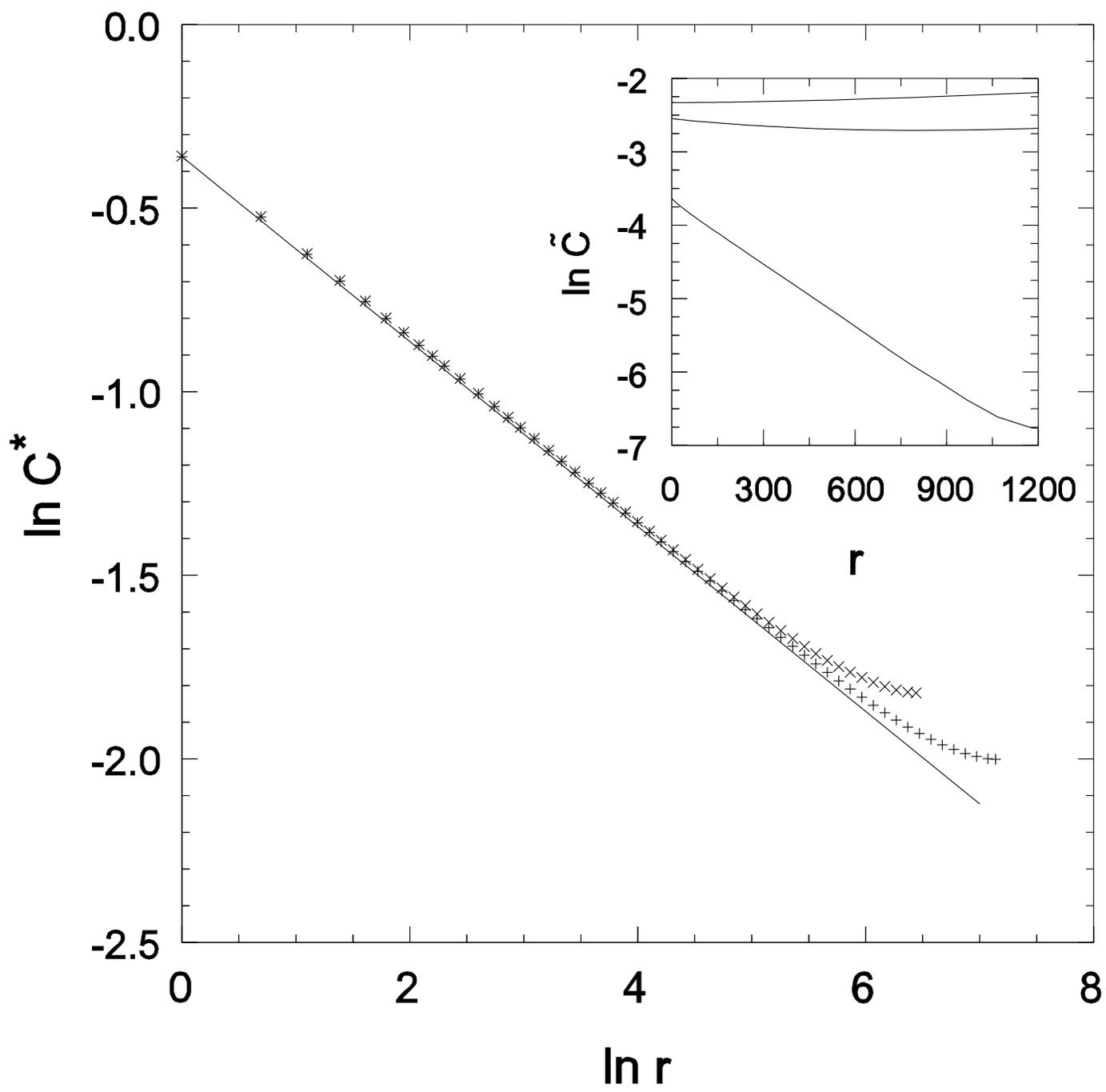
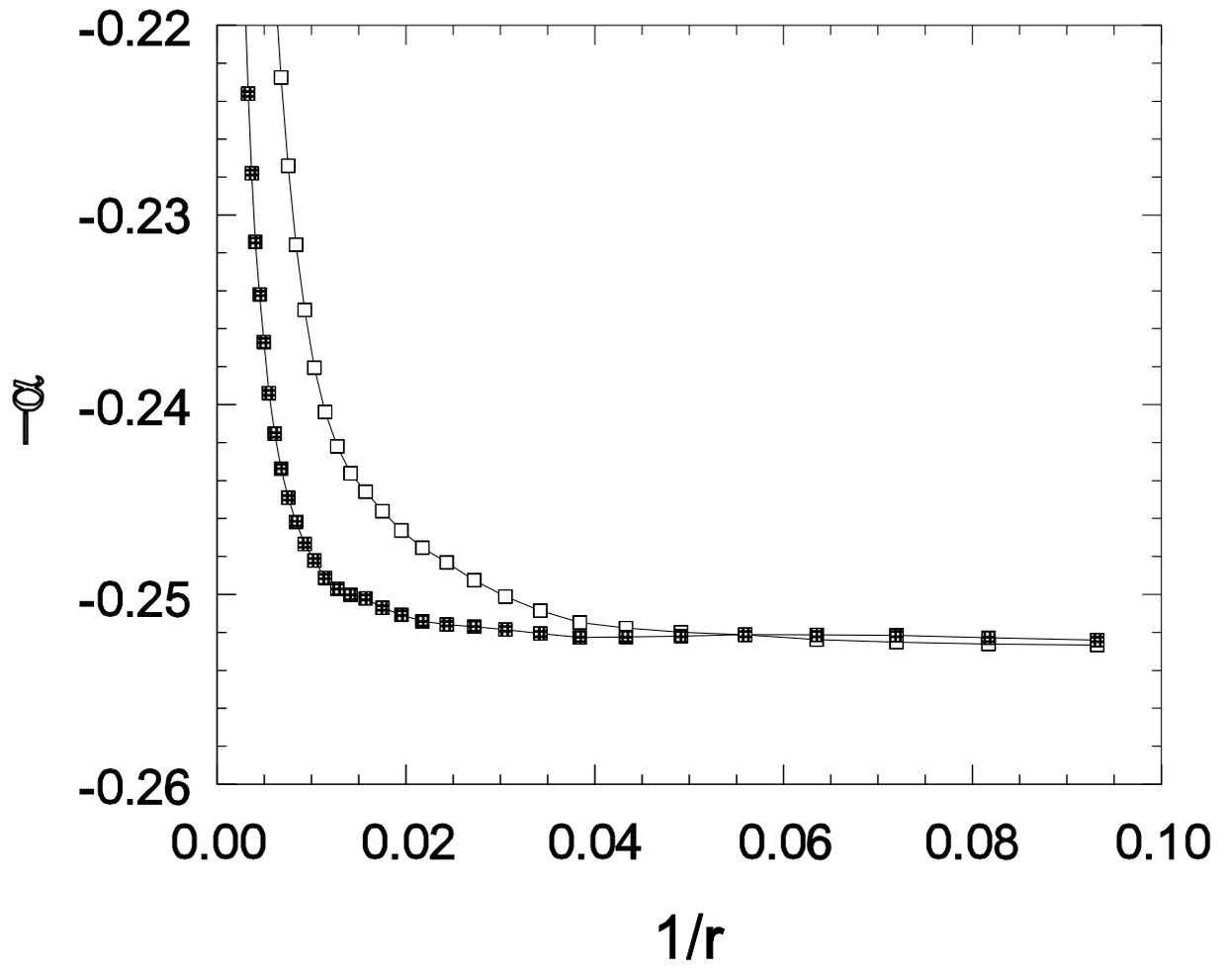
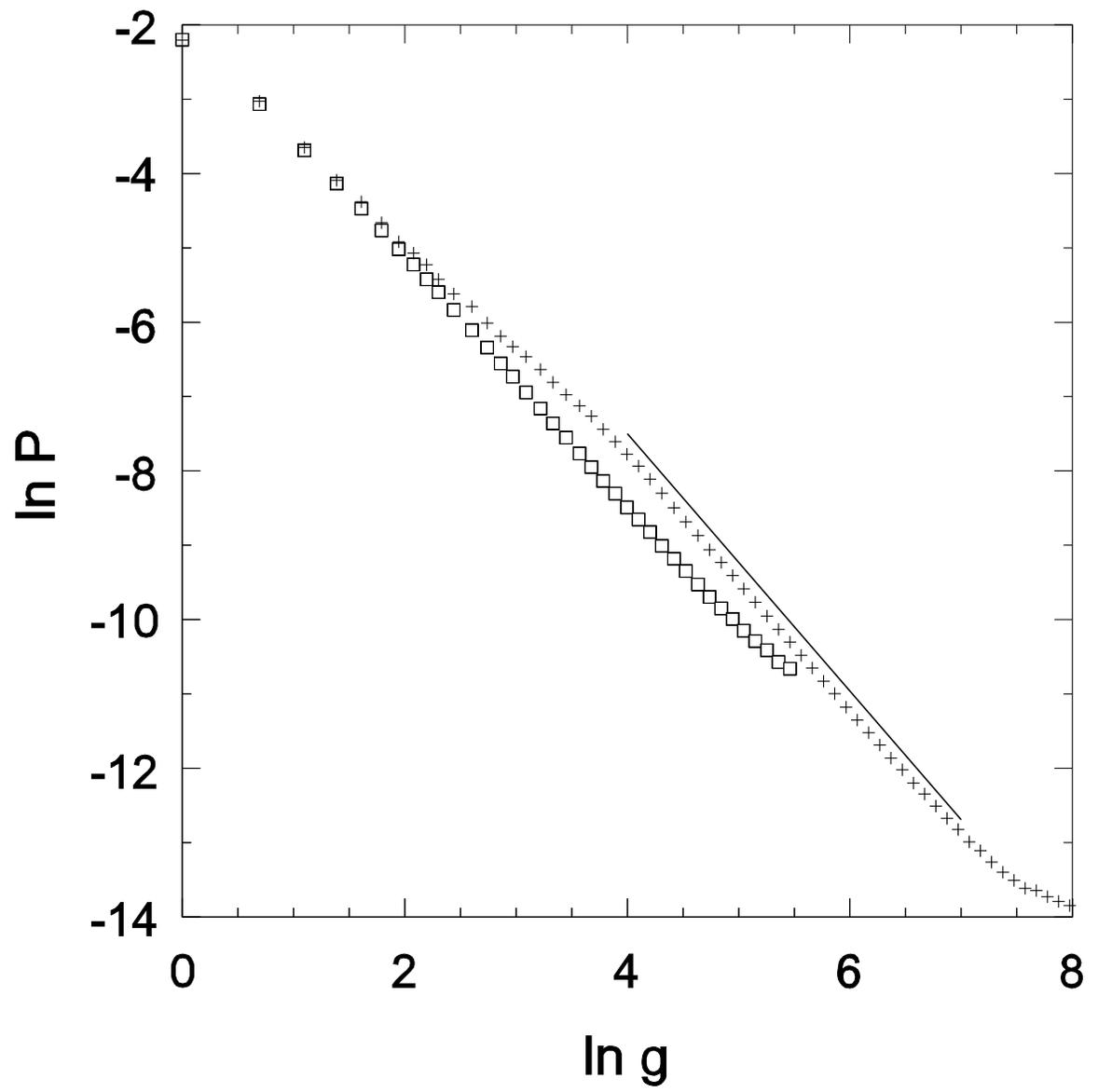
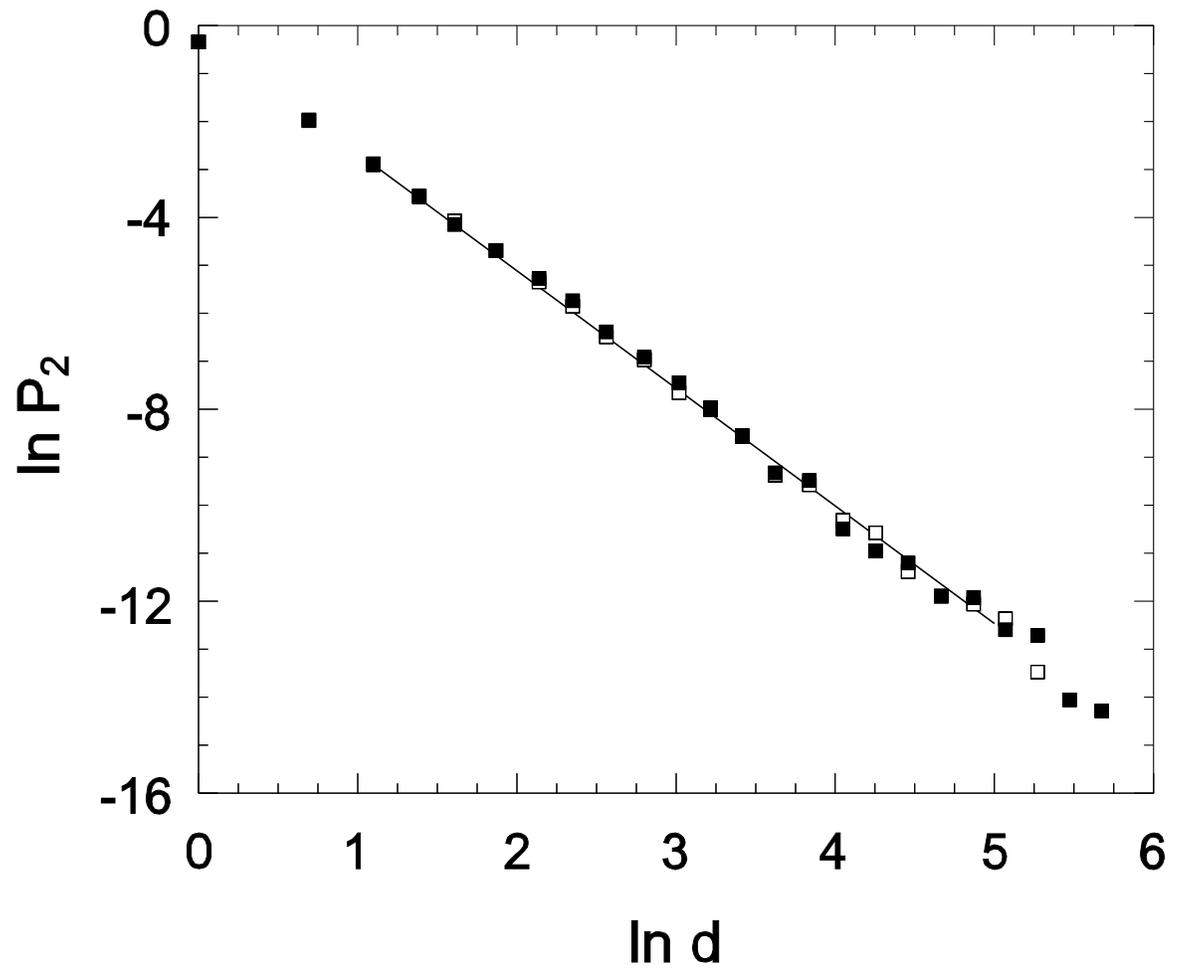
\end{document}